\documentclass[12pt]{iopart}
\expandafter\let\csname equation*\endcsname\relax
\expandafter\let\csname endequation*\endcsname\relax

\usepackage{graphicx}
\usepackage{cite}
\usepackage[colorlinks=true,allcolors=blue]{hyperref}
\usepackage{amsmath}

\newcommand{\HH}{\mathcal{H}}

\newcommand{\TBr}{\mathbf{r}}
\newcommand{\TBR}{\mathbf{R}}
\newcommand{\TBk}{\mathbf{k}}

\newcommand{\Bk}{\ensuremath{\mathbf{k}}}
\newcommand{\Br}{\ensuremath{\mathbf{r}}}
\newcommand{\Bx}{\ensuremath{\mathbf{x}}}
\newcommand{\By}{\ensuremath{\mathbf{y}}}
\newcommand{\BR}{\ensuremath{\mathbf{R}}}
\newcommand{\BA}{\ensuremath{\mathbf{A}}}
\newcommand{\Bz}{\ensuremath{\mathbf{z}}}

\newcommand{\llangle}{\ensuremath{\langle\!\langle}}
\newcommand{\rrangle}{\ensuremath{\rangle\!\rangle}}

\newcommand{\meV}{\ensuremath{\mathrm{meV}}}

\begin{document}
\title{Signatures of unconventional pairing in near-vortex electronic structure of LiFeAs}
\author{Kyungmin Lee, Mark H. Fischer, Eun-Ah Kim}
\address{Department of Physics, Cornell University, Ithaca, New York 14853, USA}
\ead{kl567@cornell.edu}

\begin{abstract}
A major question in Fe-based superconductors remains the structure of the pairing, in particular whether it is of unconventional nature.
The electronic structure near a vortex can serve as a platform for phase-sensitive measurements to answer this question.
By solving the Bogoliubov-de Gennes equations for LiFeAs, we calculate the energy-dependent local electronic structure near a vortex for different nodeless gap-structure possibilities. 
At low energies, the local density of states (LDOS) around a vortex is determined by the normal-state electronic structure.
At energies closer to the gap value, however, the LDOS can distinguish an anisotropic $s$-wave gap from a conventional isotropic $s$-wave gap.
We show within our self-consistent calculation that in addition, the local gap profile differs between a conventional and an unconventional pairing.
We explain this through admixing of a secondary order parameter within Ginzburg-Landau theory.
In-field scanning tunneling spectroscopy near a vortex can therefore be used as a real-space probe of the gap structure.
\end{abstract}

\maketitle

\section{Introduction}

The gap structure in the Fe-based superconductors and its possible unconventional nature is still a key issue in the field four years after their discovery.
In most compounds, the pairing is believed to be of the so-called $s^\pm$ type, for which the order parameter changes sign between the electron-like and the hole-like Fermi surfaces~\cite{hirschfeld:2011,stewart:2011}. Some experimental results are consistent with this prediction\cite{chen:2010,hanaguri:2010,christianson:2008,lumsden:2009,li:2009}.
However, a major difficulty in distinguishing such an unconventional pairing state from a trivial $s$-wave gap is that both states are nodeless and transform trivially under all the symmetry operations of the material's point group.
As the experimental probes that are usually used to distinguish various gap structures, such as phase-sensitive probes, are not Fermi pocket specific, an unambiguous evidence of the unconventional $s^\pm$ pairing remains evasive.

One route to accessing phase information using a phase-insensitive probe would be through vortex bound states, as a vortex introduces a spatial texture to the superconducting order parameter.
Advancements in in-field scanning tunneling spectroscopy (STS) have enabled the study of vortex bound states.
Indeed, a recent STS experiment on LiFeAs under a magnetic field has shown an intriguing energy dependence in the spatial distribution of the local density of states (LDOS) near a vortex\cite{hanaguri:2012}. The remaining question is whether the observed LDOS distribution near vortex can be instrumental in selecting one of the proposed gap structures: $s^\pm$-wave \cite{hirschfeld:2011}, $s^{++}$-wave \cite{kontani:2010, borisenko:2012}, and (spin-triplet) $p$-wave \cite{brydon:2011,haenke:2012}.
At zero bias, the LDOS shows a four-fold star shape with high-intensity `rays' along the Fe-As direction.
Similar features in NbSe$_2$~\cite{hayashi:1996} were interpreted as a sign of gap minima along this direction.
However, a quasi-classical analysis by Wang \textit{et al.}\cite{wang:2012} pointed out that  the normal-state band structure of LiFeAs -- namely a highly anisotropic hole pocket around the $\Gamma$ point -- could be producing these rays irrespective of gap structure.
By contrast, little attention has been given to the high energy LDOS distribution observed in Ref.~\cite{hanaguri:2012}: hot spots appearing at the intersection of split rays.

Motivated by these observations, we present a study of the near-vortex electronic structure and signatures of unconventional pairing therein within the Bogoliubov-de Gennes (BdG) framework.
By (non-self-consistently) imposing a gap structure and solving the BdG Hamiltonian, we first show that the isotropic $s$-wave and $s^\pm$-wave pairing result in different spatial distributions of the LDOS at energies approaching the gap value. 
In particular, we find $s^\pm$-wave pairing to yield the observed hot spots.
Then we solve the BdG equations self-consistently, and based on our results propose detecting the spatial distribution of the gap around a vortex for a more direct evidence of unconventional $s^\pm$-wave pairing.
A vortex not only suppresses the order-parameter amplitude at its core and introduces a singular point in space around which the phase of the order parameter winds, but it also induces a secondary order parameter in its vicinity 
\cite{joynt:1990,berlinsky:1995,yong:1995,ichioka:1996a,heeb:1996}.
Due to the induced secondary order parameter near the vortex, the gap recovery should show a strong angular dependence. 
Detection of such anisotropy will be an unambiguous evidence of unconventional pairing. 

The remainder of this paper is organized as follows:
In sections 2 and 3, we introduce the microscopic model and describe the Bogoliubov-de Gennes calculations, respectively.
In section 4, we present the results of the BdG calculations and discuss them within Ginzburg-Landau theory.
In section 5, we summarize our findings and remark on future directions. Throughout the paper we focus on the large hole pocket and study the single band problem.
However, we also present results from non-self-consistent BdG calculations on a five-band model in section 4, which show good agreement with observations from single-band model calculations in the energy range of our interest.

\section{Model}

\begin{figure}[tb]
  \begin{center}
    \includegraphics{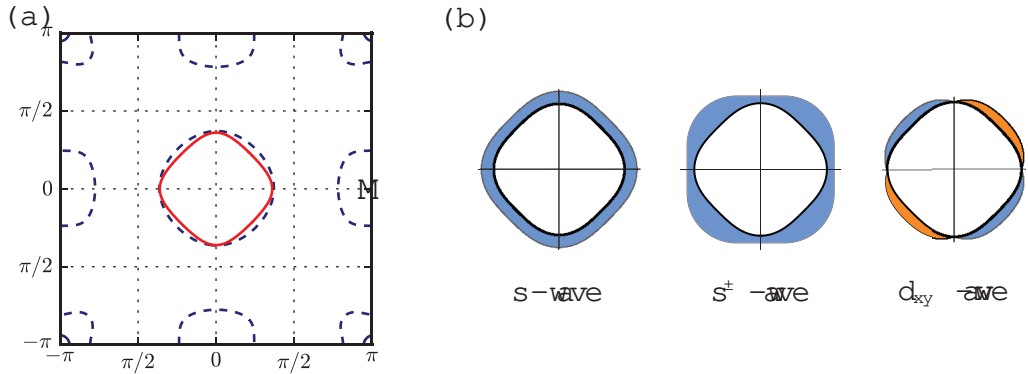}
  \end{center}
  \caption{ (a) Comparison of two tight-binding models for LiFeAs used in this paper in the 1-Fe Brillouin zone. The dashed lines indicate the Fermi surfaces of the five-band model from Ref.~\cite{eschrig:2009}.
For the most part of this work, we focus on the $\gamma$ band that is around the $\Gamma$ point, whose Fermi surface is shown as a solid line. 
  (b) Sketch of the three gap functions with $s$-, $s^\pm$-, and $d_{xy}$-wave momentum structure around the $\gamma$-band Fermi surface.
}
  \label{fig:fs}
\end{figure}

We describe LiFeAs in the superconducting state with the (mean-field) BdG Hamiltonian
\begin{align}
\HH^{\rm BdG} &=
\sum_{ij}
\Psi_{i}^{\dagger} 
\begin{pmatrix} %
  -t_{ij}         & \Delta_{ij} \\
  \Delta_{ij}^{*} & t_{ij}^{*}
\end{pmatrix}
\Psi_{j}.
\label{eq:hbdg}
\end{align}
Here, $\Psi_{i} \equiv ( c^{\phantom{\dagger}} _{i\uparrow}, c^{\dagger}_{i\downarrow} )^T$ is a Nambu spinor, and $c_{i s}$ ($c^{\dag}_{i s}$) annihilates (creates) an electron at lattice site $i$ with spin $s$ within a single-band model for the large hole pocket  around the $\Gamma$ point: the so-called $\gamma$ band.
However, Eq.~\eqref{eq:hbdg} can easily be generalized for a multi-band model.
In this paper, we focus on the single-band model for the most part since the superconducting gap is the smallest on the $\gamma$ band\cite{borisenko:2010, umezawa:2012} and hence we expect low energy physics to be dominated by this band.
Moreover, this band mainly stems from the (in-plane) $d_{xy}$ orbitals, and thus shows little $k_z$ dependence\cite{hajiri:2012}.
It is therefore a natural choice for LiFeAs.
Note that previous BdG calculations on different Fe-pnictides focused on two-band models for the $d_{xz}$ / $d_{yz}$ orbitals \cite{hu:2009,jiang:2009,zhou:2011,gao:2011,hung:2012, ye:2013}.
Our choice of the hopping matrix $t_{ij}$ is guided by the experimental observations on the $\gamma$ pocket\cite{allan:2012, borisenko:2010, umezawa:2012} to be $t=-0.25$eV for nearest-neighbor hopping, $t'=0.082$eV for next-nearest-neighbor hopping, and  $t_{ii} = \mu=0.57$eV for the chemical potential.
Figure \ref{fig:fs}(a) shows the resulting Fermi surface in solid red line. 
Though we stay within this single-band model for the self-consistent BdG studies, we have also used a five-band model for the non-self-consistent calculation with tight-binding parameters from Ref.~\cite{eschrig:2009} to test the validity of focussing on the $\gamma$ band for the energy range of our interest (see section 4.2).
Figure \ref{fig:fs}(a) shows the  Fermi surface of the five-band model in dashed lines.

The $\Delta_{ij}$ are the (bond) gap functions.
For a self-consistent solution of $\HH^{\rm BdG}$, we require the gap functions to satisfy
\begin{align}
\Delta_{ij} =& \frac{1}{2} V_{ij}
\left\langle
  c_{i\downarrow} c_{j\uparrow} +  
  c_{j\downarrow} c_{i\uparrow} 
\right\rangle ,
\label{eq:selfconsistency}
\end{align}
where $V_{ij}<0$ is the attractive interaction between sites $i$ and $j$ in the singlet channel, and $\langle \cdot \rangle$ denotes the thermal expectation value.
Restricting the interaction $V_{ij}$ to a specific form constrains the momentum structure of the gap function, since $\Delta_{ij} \neq 0$ only if $V_{ij} \neq 0$.
In the uniform case, an on-site attraction $V_{ij} = U \delta_{ij}$ leads to a spin-singlet $s$-wave gap $\Delta (\Bk) = \Delta_{s}^0$, while a next-nearest-neighbor (NNN) attraction $V_{ij} = V' \delta_{\llangle i,j\rrangle}$ allows for the singlet gap functions of  $s^\pm$ form, $\Delta (\Bk) = 4 \Delta_{s^\pm}^0 \cos k_x \cos k_y$, and $d_{xy}$ form, $\Delta (\Bk) = 4 \Delta_{d_{xy}}^0 \sin k_x \sin k_y$.
Figure~\ref{fig:fs}(b) shows sketches of these gap functions. We restrict our calculations in the following to these `pure' gap structures.
Even though the true gap function is a (symmetry-allowed) mixture of such gap functions, the dominant channel (on-site or NNN interactions) will determine whether an $s^\pm$- or an $s^{++}$-wave gap is realized in the presence of the electron pockets.

For the non-self-consistent BdG study, the vortex will be imposed through the gap-function configuration of 
\begin{align}
  \Delta_{ij} =& \Delta^{0} \tanh ( |\Br_{ij}|/ \xi ) e^{i \theta_{ij}},
	\label{eq:nonsc_gap}
\end{align}
where the vector $\Br_{ij}$ points to the midpoint of sites $i$ and $j$, and $\theta_{ij}$ is its azimuthal angle measured from the Fe-Fe direction.
This corresponds to a single vortex located at the origin suppressing locally the order-parameter amplitude. In addition, the order-parameter phase winds around the vortex core.

For the self-consistent BdG study, we induce the vortices by applying a magnetic field $H \hat{\Bz}$.
Assuming minimal coupling between an electron and the field, the hopping between sites $i$ and $j$ acquires a Peierls phase
\begin{align}
  t_{ij} \; \longrightarrow \; & t_{ij} e^{i \varphi( \Br_{i}, \Br_{j} )},&
\varphi(\Br_{i}, \Br_{j}) \equiv& -\frac{\pi}{\Phi_{0}} \int_{\Br_{j}}^{\Br_{i}} \BA(\Br)
  \cdot d \Br, 
  \label{eq:kij}
\end{align}
where $\Phi_{0}= h / 2 e$ is the magnetic fluxoid and $\Br_{i}$ is the vector pointing to the site $i$.
We assume a uniform magnetic field $H$ and write the vector potential in the Landau gauge $\BA (\Br)= -Hy \hat{\mathbf{x}}$.
From the self-consistent solution $\Delta_{ij}$, we can define local gap order parameters of different symmetries.
For an on-site interaction, the local $s$-wave order parameter is defined as $\Delta_{s}(\Br) = \Delta_{\Br, \Br}$.
Note that from here on, we use $\Br$ without any site index to denote both a lattice site and the vector pointing to it in units of the lattice constant $a_{0}$.
With NNN interaction, we define local order parameters of $s^\pm$ form
\begin{align}
 \Delta_{s^\pm}(\Br) = \frac{1}{4}
  [ \widetilde{\Delta}_{\Br + (1, 1), \Br} +
  \widetilde{\Delta}_{\Br + (1, -1), \Br}
  + \widetilde{\Delta}_{\Br + (-1, -1), \Br} +
  \widetilde{\Delta}_{\Br + (-1, 1), \Br} ]
  \label{eq:gap_local_spm}
\end{align}
and
$d_{xy}$ form
\begin{align}
  \Delta_{d_{xy}}(\Br) = \frac{1}{4}
  [ \widetilde{\Delta}_{\Br + (1, 1), \Br} -
  \widetilde{\Delta}_{\Br + (1, -1), \Br}
  + \widetilde{\Delta}_{\Br + (-1, -1), \Br} -
  \widetilde{\Delta}_{\Br + (-1, 1), \Br} ],
  \label{eq:gap_local_dxy}
\end{align}
where $\widetilde{\Delta}_{\Br \Br'}\equiv
  \Delta_{\Br \Br'} 
  \exp[- i \varphi (\Br, \Br')]$
ensures that order parameters of different symmetries do not mix under magnetic translations.
Note that for the uniform case, $\Delta_{s}(\Br) = \Delta_{s}^{0}$, $\Delta_{s^\pm}(\Br) = \Delta_{s^\pm}^{0}$, and $\Delta_{d_{xy}}(\Br) = \Delta_{d_{xy}}^{0}$ as defined above.

\section{Method}
In this section, we elaborate on our two approaches to solve the BdG equations and obtain the LDOS near a vortex.
For both, diagonalizing the Hamiltonian $\HH^{\mathrm{BdG}}$ in Eq.~\eqref{eq:hbdg} for a system of size $(N_x, N_y)$ is computationally the most expensive part.


\subsection{Non-Self-Consistent Approach}

For the non-self-consistent calculation, we impose a gap function in the form given by Eq.~\eqref{eq:nonsc_gap} and find the low lying eigenvalues and eigenstates of $\HH^{\mathrm{BdG}}$ using the Lanczos algorithm\footnote{We suppress low energy states from forming at the boundary by imposing an on-site potential of 10 eV to the sites at the boundary.}.
The LDOS can be calculated from the eigenenergies $E^{n}$ and eigenstates $[u^n (\Br), v^n (\Br)]$ as
\begin{align}
  N(\Br, E)  
  =\! \sum_{n}
    | u^n (\Br) |^2 \delta( E - E^n ) +
    | v^n (\Br) |^2 \delta( E + E^n )
.
\label{eq:nonsc_ldos}
\end{align}
Since we are not interested in the absolute value of the LDOS but rather in the spatial profile at a given energy, we normalize the LDOS such that for a given energy $E$, the maximum value of $N(\TBr, E)$ is unity.


\subsection{Self-Consistent Approach }

For the self-consistent calculation, we assume initial gap functions and use the eigenvalues and eigenvectors of Eq.~\eqref{eq:hbdg} to calculate the gap functions given by Eq.~\eqref{eq:selfconsistency}.
We proceed iteratively until self-consistency is achieved.
In diagonalizing $\HH^{\mathrm{BdG}}$, we can no longer make use of the crystal momentum basis to simplify the problem since the Peierls phase factor prevents the kinetic part of the Hamiltonian from commuting with the ordinary lattice translation operator $T_{\BR}$.
However, the kinetic part commutes with the magnetic translation operator
\begin{align}
\hat{T}_{\BR}
&\equiv
 e^{ -i \frac{\pi}{\Phi_{0}} \BA (\BR) \cdot \Br } T_{\BR}
\end{align}
for a magnetic lattice vector $\BR$ whose unit cell contains two magnetic fluxoids.

The pairing term in general does not commute with $\hat{T}_{\TBR}$.
Nevertheless, when vortices form a lattice, $\hat{T}_{\TBR}$ commutes with the pairing term when $\TBR$ is a vector of a vortex sublattice containing every other vortex.
Since we focus on the electronic structure near a single vortex, we expect the shape of the vortex lattice to have little influence on our results.
Therefore, we make an arbitrary choice for its primitive vectors to be $L_x \hat{\Bx}$ and $L_y \hat{\By}$, such that $\TBR$ forms a rectangular lattice $\TBR = ( m_x L_x , m_y L_y )$, where $m_\alpha = 0 \cdots M_\alpha - 1$ and $M_\alpha \equiv N_\alpha / L_\alpha$\footnote{This choice yields an oblique vortex lattice, since there are two vortices in each (rectangular) magnetic unit cell, trying to form a triangular vortex lattice as a self-consistent solution.}.
Note that periodic boundary conditions in the Landau gauge $\BA(\Br) = -H y \hat{\Bx}$ require the total magnetic flux through the system to be an integer multiple of $2 \Phi_0 N_x$.
In addition, one magnetic unit cell contains a magnetic flux of $2 \Phi_0$, i.e. $H = 2 \Phi_0 / L_x L_y$.
We satisfy these two requirements by choosing $M_x = L_y , M_y = L_x$.

Working with the magnetic Bloch states
\begin{align}
\Psi_{\TBk} (\Br) &= \sum_{\TBR} e^{-i \TBk \cdot \TBR} \; \hat{T}_{\TBR} \Psi (\Br) \hat{T}_{\TBR}^{-1} 
\label{eq:mbloch}
\end{align}
allows us to block diagonalize the Hamiltonian
\begin{align}
  \HH^{BdG} &= \frac{1}{M_{x} M_{y}} 
  \sum_{\TBk}\sum_{\TBr, \TBr' }
  \Psi_{\TBk}^\dagger ( \TBr ) 
  H_{\TBk} (\TBr, \TBr')
  \Psi_{\TBk} ( \TBr' )
.
\label{eq:hmbloch}
\end{align}
The indices $\TBk$ and $\TBr$ from here on are defined in the magnetic Brillouin zone and magnetic unit cell, respectively, that is
\begin{subequations}
\begin{align}
\TBk &= \left( 2\pi \frac{m_{x}}{L_{x}M_{x}},2\pi \frac{m_{y}}{L_{y}M_{y}} \right),
&
m_{\alpha} &= 0 \cdots M_{\alpha}-1,\\
\TBr&= \left( \ell_{x}, \ell_{y} \right), &
\ell_{\alpha} &= 0 \cdots L_{\alpha}-1.
\end{align}
\end{subequations}
By diagonalizing the matrices $H_{\TBk}$ of dimension $2 L_{x} L_{y} \times 2 L_x L_y$ in Eq.~\eqref{eq:hmbloch}, we can compute the eigenstates and eigenenergies of $\HH^{BdG}$.
These are then used to calculate $\Delta_{ij}$ with Eq.~\eqref{eq:selfconsistency} closing the self-consistency cycle. Finally, we use the self-consistent solution $\Delta_{ij}$ to calculate the local order parameters of $s$-, $s^\pm$- and $d_{xy}$-wave symmetry and also the LDOS of the electronic degrees of freedom, as defined in Eq.~\eqref{eq:nonsc_ldos}.

\section{Results}
\subsection{Non-Self-Consistent Approach on Single Band Model}
\begin{figure*}[tb]
  \begin{center}%
  \begin{minipage}{100pt}
	\includegraphics[height=115pt]{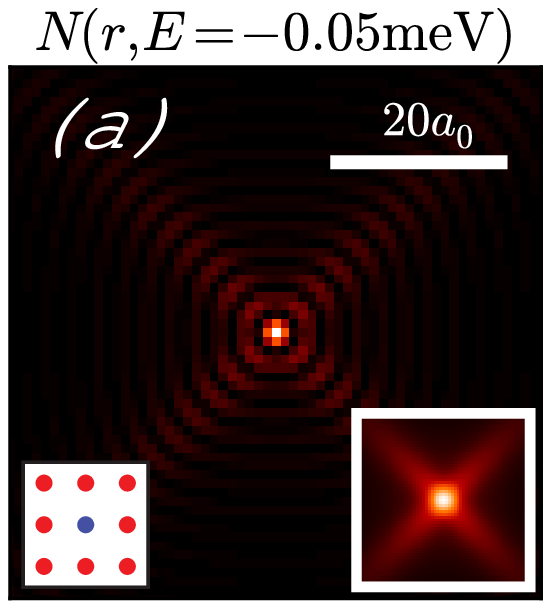}\\
    \includegraphics[height=115pt]{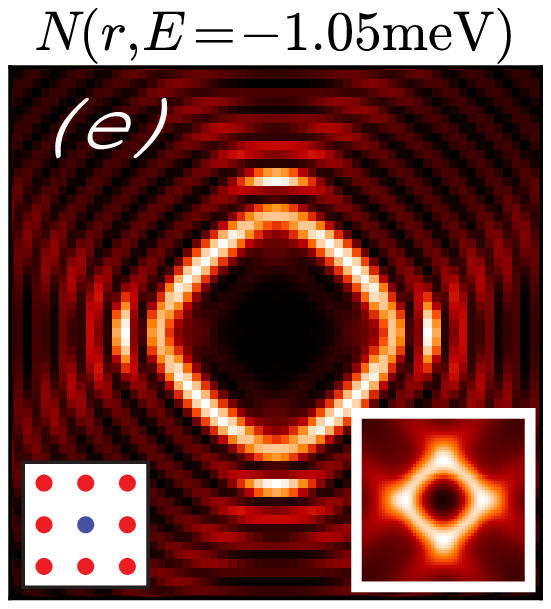}
  \end{minipage}%
  \begin{minipage}{120pt}
  	\includegraphics[height=115pt]{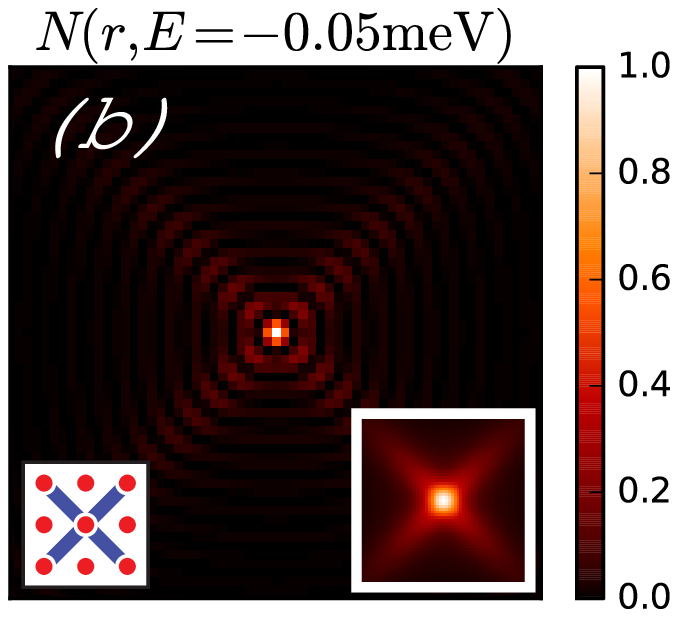} \\
    \includegraphics[height=115pt]{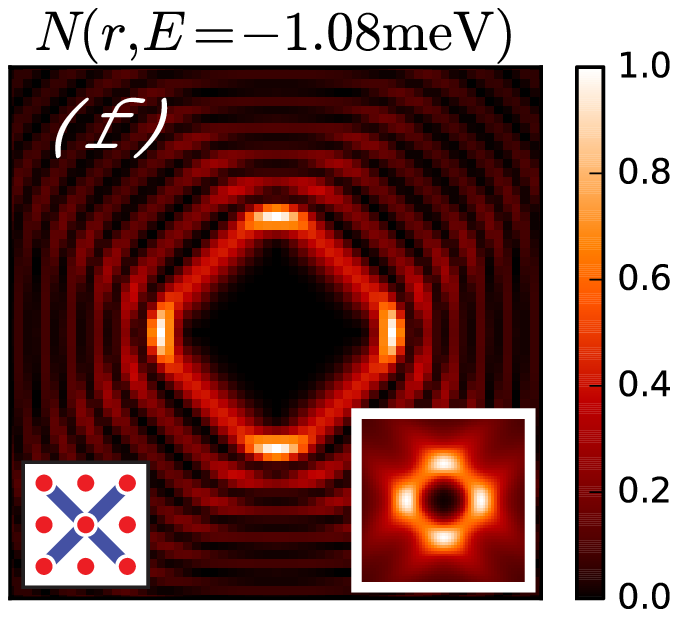} 
  \end{minipage}%
  \begin{minipage}{110pt}
    \includegraphics[height=115pt]{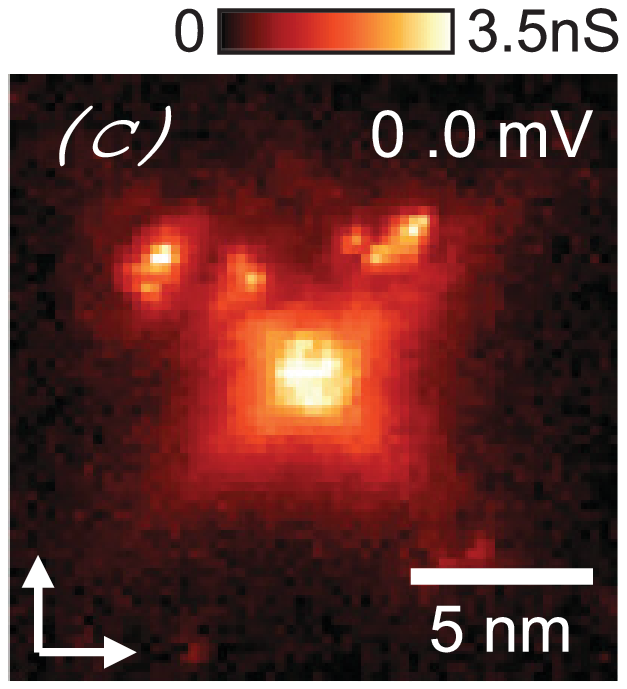} \\
 	\includegraphics[height=115pt]{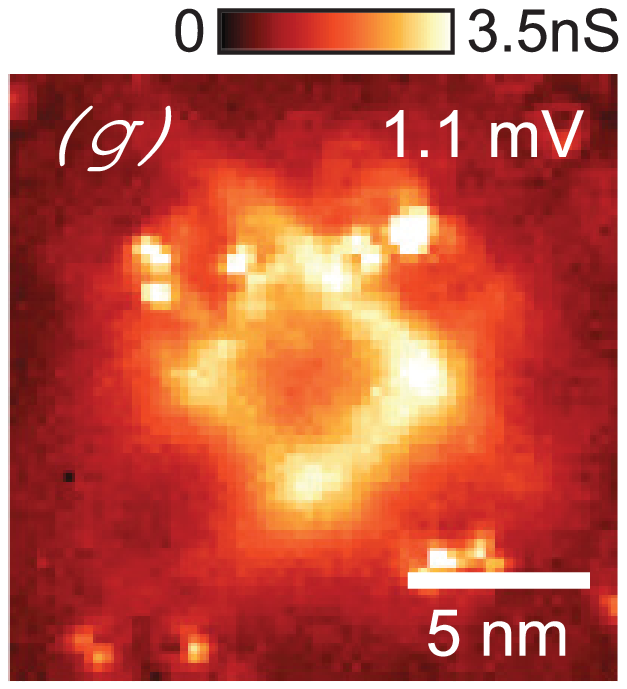} 
  \end{minipage}%
  \begin{minipage}{120pt}  
    \includegraphics[height=115pt]{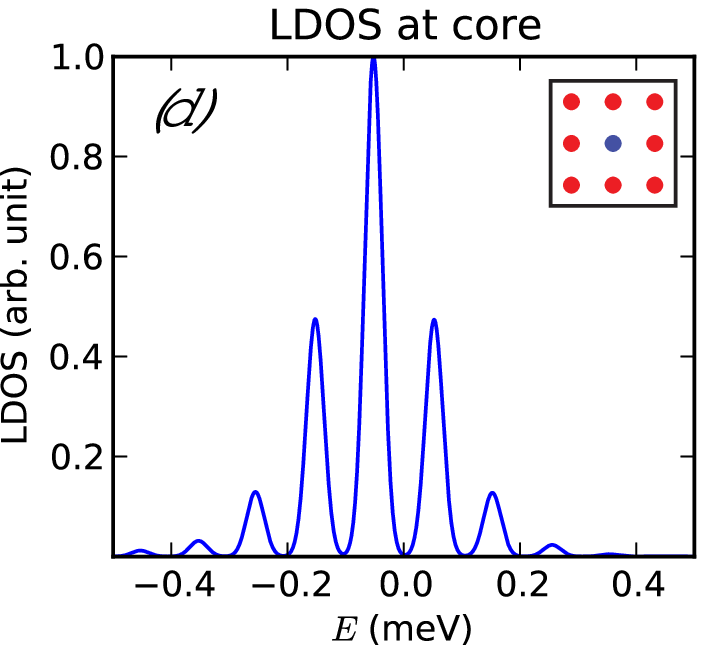}
    \includegraphics[height=115pt]{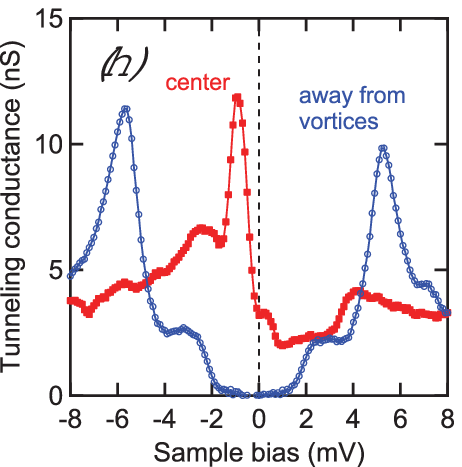} 
  \end{minipage}%
  \end{center}
  \caption{\label{fig:nonsc_ldos}
Local density of states near a vortex for the non-self-consistent calculation with the gap function given by Eq.~\eqref{eq:nonsc_gap}.
The value $N(\Br,E)$ has been normalized such that the maximum value in each map is unity.
(a) shows the LDOS at the lowest bound state energy with on-site pairing with $\Delta^0=3\meV$, and (e) is at higher energy.
(b) and (f) are with NNN pairing with $\Delta^0=1.5\meV$.
The left insets in (a),(b),(e) and (f) indicate the local structure of the pairing, and the right insets are LDOS after gaussian filtering ($\sigma = 3 a_0$) reducing spatial resolution for better comparison with experiment\cite{hanaguri:2012}.
(c) and (g) are the near-vortex LDOS maps observed in Ref.~\cite{hanaguri:2012}.
(d) is the LDOS as a function of energy at the vortex core for the on-site pairing, Gaussian-filtered in both energy ($\sigma = 0.15\meV$) and position ($\sigma = a_{0}$).
(h) shows the experimental tunneling spectra from Ref.~\cite{hanaguri:2012} for comparison.
}
\end{figure*}

Figure \ref{fig:nonsc_ldos} shows the near-vortex LDOS calculated by diagonalizing $\HH^{\mathrm{BdG}}$ of Eq.~\eqref{eq:hbdg} with fixed gap functions as given by Eq.~\eqref{eq:nonsc_gap} on a system of dimension $(N_x, N_y) = (301,301)$.
We choose realistic values of the parameters for the coherence length $\xi = 16.4 a_{0}$\cite{lee:2010,kurita:2011}, as well as gap values $\Delta_{s}^{0} = 3\meV$ for on-site pairing and $\Delta_{s^\pm}^{0} = 1.5 \meV$ for NNN pairing\cite{borisenko:2010, umezawa:2012, allan:2012}.

We can interpret the vortex bound states in this non-self-consistent BdG calculation as bound states in a potential well given by Eq.~\eqref{eq:nonsc_gap}, where only states around the normal-state Fermi surface constitute the bound states.
There are then two sources of anisotropy: anisotropic, quasi-one-dimensional low-energy properties of the normal state, and an anisotropic gap, both defined in the momentum space.
The geometric distribution of LDOS will be dominated by one or the other source of anisotropy at different energies. 

At low energies, the normal state properties dominate the distribution of LDOS [Figs.~\ref{fig:nonsc_ldos}(a) and (b)]. 
Hence irrespective of pairing structure, the bound state is located at the center of the potential well.
Since the Bloch states making up this bound state have two main velocities due to the quasi-one-dimensional parts of the Fermi surface, the bound state mainly extends in these two directions out of the well, resulting in the rays in Figs.~\ref{fig:nonsc_ldos}(a) and (b).
The gap is suppressed near the vortex center, and its anisotropy is of little importance.
Hence the flat (quasi-one-dimensional) parts of the electronic structure in Fig.~\ref{fig:fs}(a) (solid line) dominate over the small anisotropy of the $s^{\pm}$ gap [see Fig.~\ref{fig:fs}(b)].
For a better comparison with experiment, we present results of reduced spatial resolution by gaussian filtering ($\sigma=3a_{0}$) in the insets. The low resolution result is consistent with results of the quasi-classical analysis by Wang \textit{et al.}~\cite{wang:2012} and in good agreement with experiment shown in Fig.~\ref{fig:nonsc_ldos}(c).

At higher energies on the other hand, the bound state is located away from the vortex core.
The quasi-one-dimensionality of the Fermi surface allows for localization in one direction and extension in the other.
This leads to a square-like inner ring in the LDOS for both pairings [Figs.~\ref{fig:nonsc_ldos}(e) and (f)].
The difference, however, results from the anisotropy of the gap function.
While the isotropic $s$-wave gap is analogous to a potential that is independent of momentum, the anisotropic gap is one for which different states around the Fermi surface experience different potentials depending on their momenta.
With the gap function of $s^\pm$ form, the quasi-one-dimensional portion of the Fermi surface experiences a stronger trap potential, leading to a suppression of its contribution to the bound-state wave function.
As a result, the bound state exhibits pronounced isolated segments, `hot spots,' within the inner ring that point in the Fe-Fe direction, as shown in Fig.~\ref{fig:nonsc_ldos}(f).
We again gaussian filter the images and show them in the insets. Note the 'hot spot' are robust and even more pronounced in the low resolution insets in Fig.~\ref{fig:nonsc_ldos}(f) in good agreement with the experimental data Fig.~\ref{fig:nonsc_ldos}(g).

We now turn to the LDOS at the core of the vortex and its particle-hole asymmetry.
This turns out to be largely insensitive to anisotropy of pairing.
The LDOS at the core of the vortex for the on-site pairing shown in Fig~\ref{fig:nonsc_ldos}(d) exhibits particle-hole asymmetry with the highest peak at negative energy.
Such asymmetry appears in the so-called `quantum-limit' vortex bound state\cite{caroli:1964}, whose highest LDOS peak is at energy $\Delta^2 / 2 E_F$ above(below) the Fermi energy for an electron(hole)-like band, where $E_F$ is the energy difference between the Fermi energy and the bottom(top) of the band.
The energy of the LDOS peak being $0.05\meV$ below the Fermi energy is expected given $E_F = 98 \meV$ and $\Delta = 3\meV$ within our input bandstructure.
Though similar particle-hole asymmetry has been observed in Ref.~\cite{hanaguri:2012} [see Fig.~\ref{fig:nonsc_ldos}(h)] the energy at which the peak was observed suggests that other hole pockets with larger gap values may be responsible.

\subsection{Non-Self-Consistent Approach on Five Band Model}

\begin{figure*}[tb]
  \begin{center}
     \includegraphics[height=115pt]{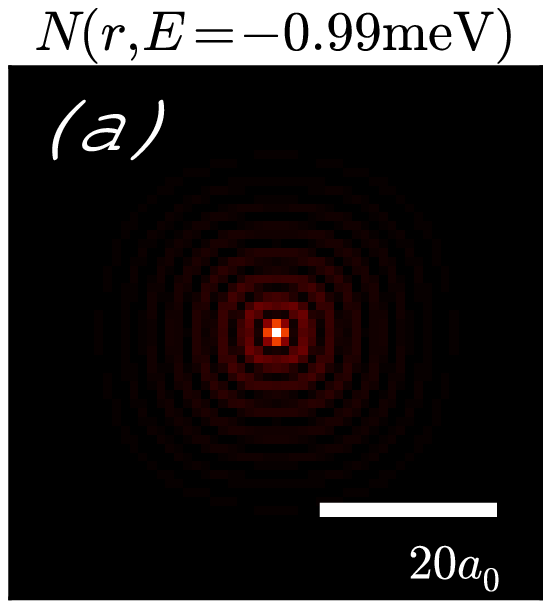}
     \includegraphics[height=115pt]{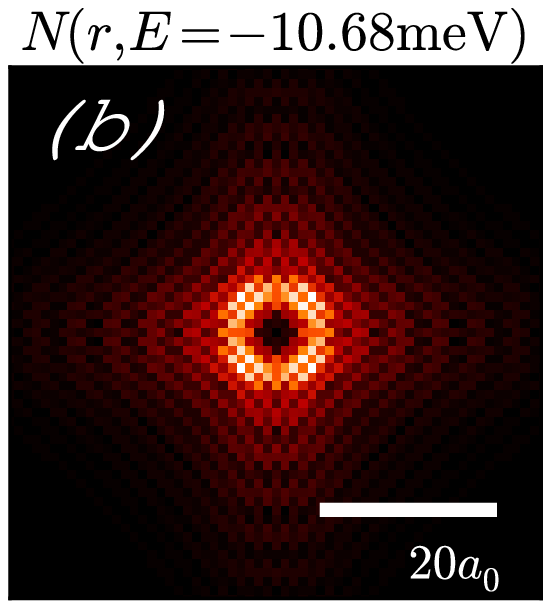}
     \includegraphics[height=115pt]{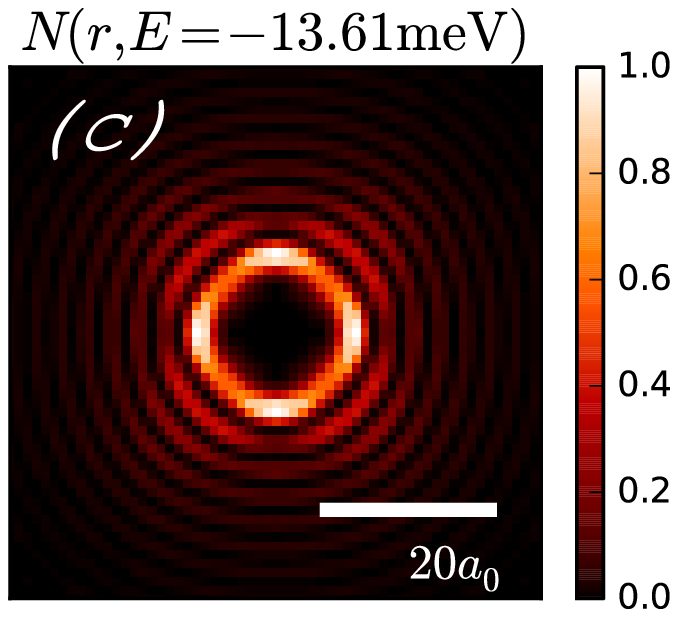}
  \caption{\label{fig:nonsc5b_ldos}
  LDOS near a vortex from the non-self-consistent calculation with the five-band model from Ref.~\cite{eschrig:2009} and NNN pairing of $\Delta^0_{s^\pm} = 15 \meV$,
  (a) at the lowest bound state energy,
  (b) at an energy where the electron-band contribution dominates, and
  (c) at an energy where the $\gamma$-band contribution dominates.
  } 
  \end{center}
\end{figure*}

Now, we check whether the single-band model is sufficient to describe vortex bound states within the energy range of interest.
A simple insight can be gained by treating each band independently and estimating the energy of its lowest bound state to be $\Delta^2/2 E_F$ following Caroli \textit{et al.}\cite{caroli:1964} for the gap size $\Delta$ and the Fermi energy $E_F$ specific to each band.
Using measured Fermi energies and gap parameters\cite{borisenko:2010, borisenko:2012, umezawa:2012,allan:2012}, we estimate the energies of the lowest bound states of the $\gamma$ pocket and the electron pockets to be of the same order. 
However, the lowest bound state energies of the two smaller hole pockets are an order of magnitude larger.
This rough estimate implies that the LDOS within the energy below $1 \meV$ should be dominated by bound states coming from the  $\gamma$ band and those coming from the two electron bands. 
If indeed each bound state comes from a single band, we expect to find bound states with LDOS distribution resembling what we predicted in section 4.1.

For concreteness, we carry out a non-self-consistent BdG calculation using the band structure given by Ref.~\cite{eschrig:2009} with five bands. 
Unfortunately, the $\gamma$-pocket Fermi surface of this band structure [dashed line in Fig 1(a)] is far more isotropic compared to what has been measured in Ref.\cite{allan:2012} and guided the band structure we use in the rest of this paper. 
Hence we do not expect as pronounced `ray' features at low energies compared to what is shown in Fig.~\ref{fig:nonsc_ldos} from our (single-band) calculations and experiment. 
Another issue we face with a five-band calculation is the limitation on the accessible system size. 
For a system of size $(101,101)$, we impose NNN pairing that is trivial in the orbital space having magnitude $\Delta_{s^\pm} = 15 \meV$ in order to fit the vortex bound states within the system and minimize the boundary effect.
As in the single-band calculation, we create a vortex at the center of the form given in Eq.~\eqref{eq:nonsc_gap}, however with $\xi = 10 a_0$.
Figure~\ref{fig:nonsc5b_ldos} shows the resulting LDOS at different bound state energies.
At the lowest energy there is no clear sign of `rays' though a small amount of anisotropy is still present, as expected from the smaller $\gamma$-band anisotropy [see Fig.~\ref{fig:nonsc5b_ldos}(a)].
Figures~\ref{fig:nonsc5b_ldos}(b) and (c) show typical LDOS images of vortex bound states at higher energies. Figure \ref{fig:nonsc5b_ldos}(b) looks very different from the LDOS distribution obtained in section 4.1 and we hence assign the corresponding bound state to the electron pockets.
However, the LDOS shown in Fig.~\ref{fig:nonsc5b_ldos}(c) shows the same `hot spots' as obtained within our single-band calculation and shown in Fig.~\ref{fig:nonsc_ldos}(f).
Focussing on the $\gamma$ band should thus suffice to capture the features observed in Ref.~\cite{hanaguri:2012}.

\subsection{Self-Consistent Approach}

\begin{figure}[tb]
  \begin{center}%
\includegraphics[height=115pt]{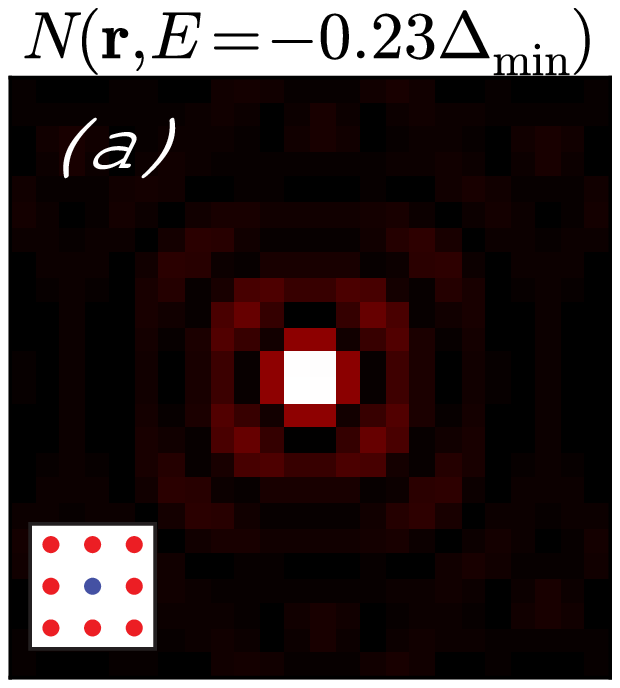}%
\includegraphics[height=115pt]{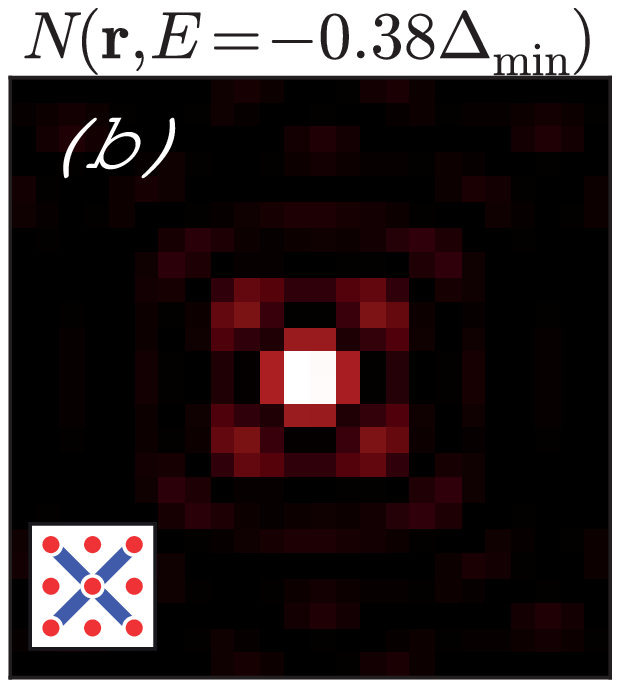}%
\includegraphics[height=115pt]{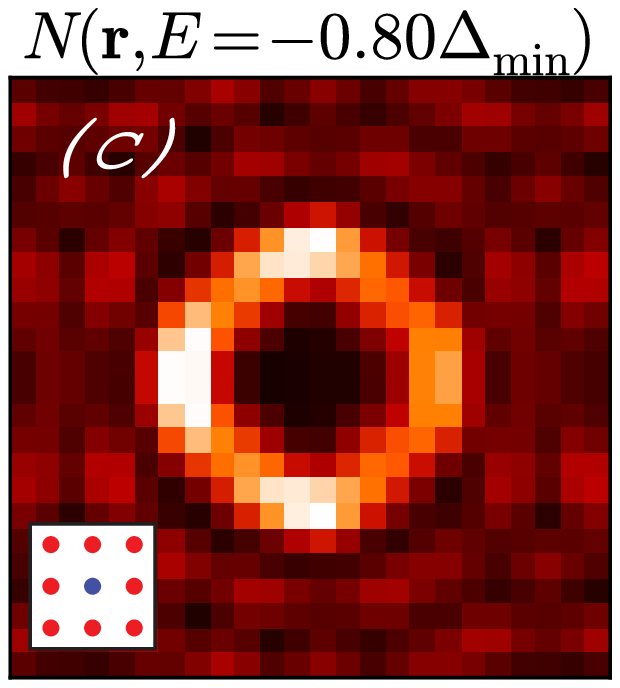}%
\includegraphics[height=115pt]{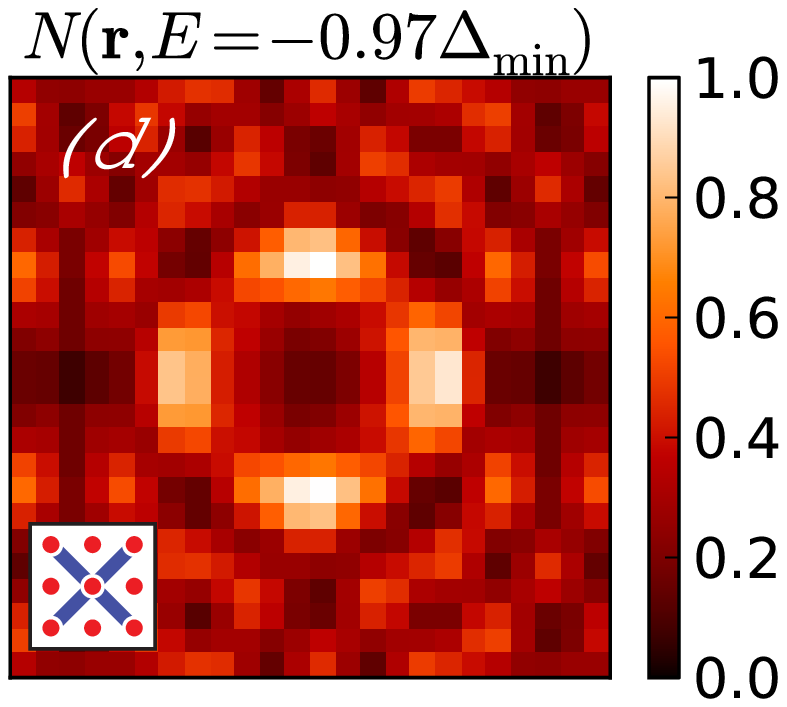}%
\end{center}%
  \caption{%
    \label{fig:sc_ldos}Local density of states near a vortex within our self-consistent calculation.
Again, $N(\Br,E)$ is normalized within each image.
(a) is the LDOS at the lowest bound state energy with on-site attraction $U=- 0.35 \mathrm{eV}$, and (c) is at higher energy.
(b) and (d) are with NNN attraction $V'=-0.3 \mathrm{eV}$.
The inset in each figure represents the local attractive interaction in the singlet pairing channel.
}%
\end{figure}

Figure \ref{fig:sc_ldos} shows the results from the (single-band) self-consistent calculation.
We compare two pairing interactions -- on-site attraction $U=-0.35\mathrm{eV}$, and NNN attraction $V'=-0.3\mathrm{eV}$ -- for a system with magnetic unit cell of dimensions $( L_x, L_y) = (19,38)$.
This corresponds to a full lattice size of $(N_x, N_y) = (38 \times 19, 19 \times 38)$.
In zero field, the two cases lead to a uniform superconducting gap of $\Delta_{s}^{0} = 27\mathrm{meV}$ and $\Delta_{s^\pm}^{0} = 10 \mathrm{meV}$, respectively.
We have chosen $U$ and $V'$ such that the coherence length $\xi \propto \Delta^{-1}$ is small compared to the inter-vortex spacing.
This allows us to focus on a nearly isolated vortex within the computationally feasible size of the magnetic unit cell.
Although the resulting gap values are an order of magnitude larger than what is known experimentally, this should not affect the validity of the results in a qualitative manner.
Both at low energy and at higher energy close to the gap value, we observe features that qualitatively agree with the results obtained in the previous section.

\begin{figure}[tb]
  \begin{center}
    \includegraphics[height=115pt]{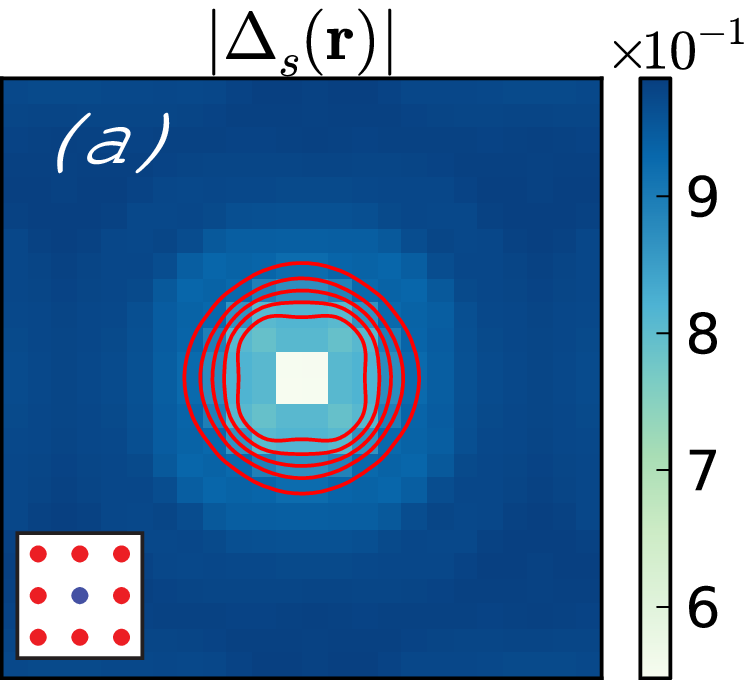}
    \includegraphics[height=115pt]{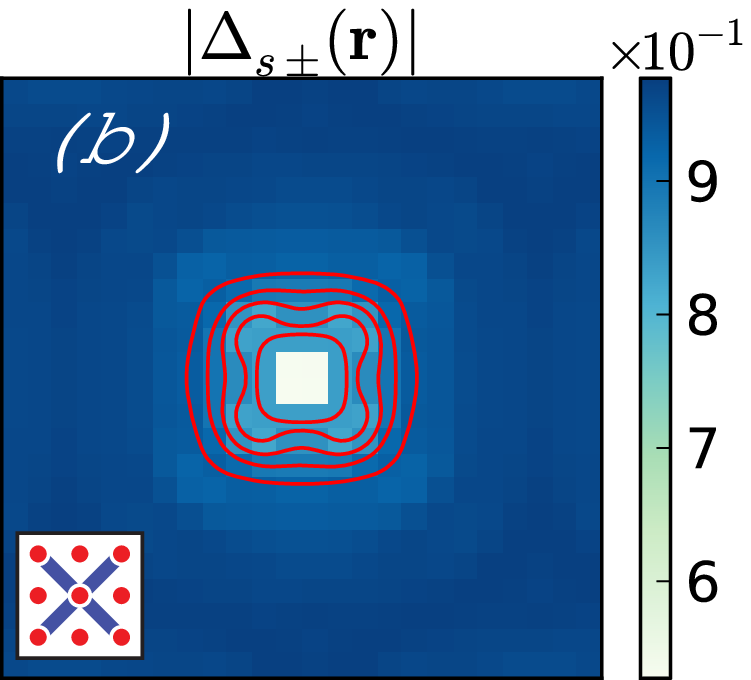}
    \includegraphics[height=115pt]{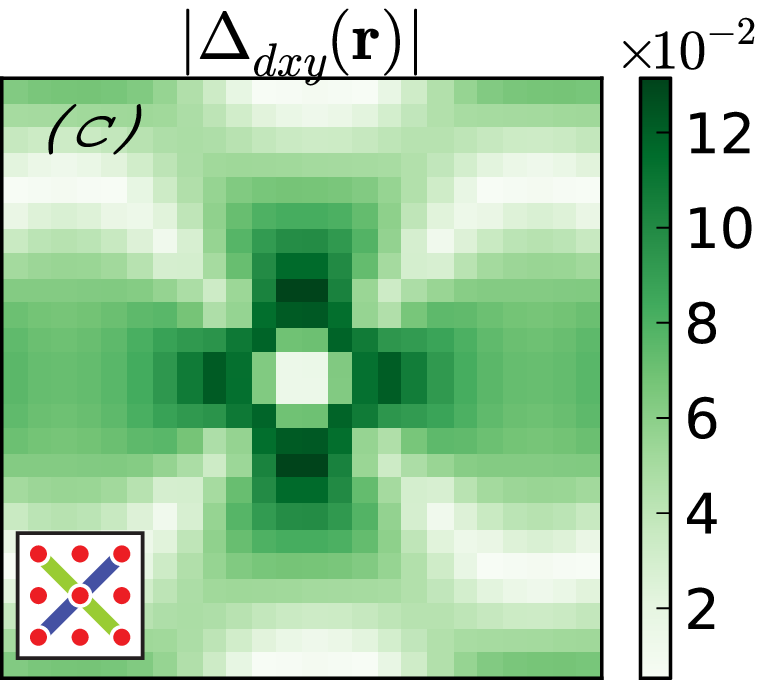}
  \end{center}
  \caption{    \label{fig:gapmap}
Spatial distribution of different symmetry components of order parameters. 
(a) $\Delta_{s}(\Br)$ for on-site attraction $U=-0.35\mathrm{eV}$.
(b) $\Delta_{s^\pm}(\Br)$ and (c) $\Delta_{d_{xy}}(\Br)$ for NNN pairing $V'=-0.3\mathrm{eV}$.
The values have been normalized by the value of the dominant order parameter in the absence of magnetic field for each case: $\Delta_{s}^{0}$ for (a), and $\Delta_{s^\pm}^{0}$ for (b), (c).
The equal-amplitude contours in red go from 0.825 for the innermost to 0.925 for the outermost contours (after normalization) with equal intervals between the contours in between.
The insets again indicate the structure of the local order parameter.
Note that the color-scale for $\Delta_{d_{xy}}(\Br)$ is much smaller than for $\Delta_{s}(\Br)$.
}
\end{figure}

The self-consistent calculation also allows us to study the local order parameters of a given structure near a vortex.
Unlike for the on-site attraction, where $\Delta_s(\Br)$ is the only allowed gap function, order parameters of different symmetries can mix near a vortex for NNN attraction.
A near-vortex map of $\Delta_{s}(\Br)$ for on-site pairing shown in Fig.~\ref{fig:gapmap}(a) indeed shows almost isotropic healing of the order parameter away from the vortex core.
However, for the NNN attraction which leads to uniform $s^\pm$-wave pairing in zero-field, the secondary order parameter $\Delta_{d_{xy}}(\Br)$ is induced near the vortex.
Coupling between this secondary order parameter and the primary $\Delta_{s^\pm}(\Br)$ leads to a strong angular variation of both components as can be seen in Figs.~\ref{fig:gapmap}(b) and (c).

To gain further insight into the admixing of a secondary order parameter near a vortex for the anisotropic pairing, we analyze the Ginzburg-Landau free-energy density.
The free-energy density for $s$-wave and $d$-wave order parameters reads
\begin{align}
  f =& 
  \alpha_{s} |s|^{2}
  + \alpha_{d} |d|^{2}
  + \beta_{1} |s|^{4}
  + \beta_{2} |d|^{4}
  + \beta_{3} |s|^{2} |d|^{2} \nonumber \\
  &
  + \beta_{4} ( s^{*2} d^{2} + \mathrm{c.c.})
  + \gamma_{s} | \vec{D} s |^{2}  
  + \gamma_{d} | \vec{D} d |^{2} \nonumber\\
  &
  + \gamma_{v} ( D_{x} s D_{y} d^{*} + D_{y} s D_{x} d^{*} + \mathrm{c.c.}),
  \label{eq:glfe}
\end{align}
where $s$ and $d$ are shorthands for $s(\Br)$ and $d(\Br)$, the order-parameter fields for the $s^{\pm}$ and $d_{xy}$ gaps, respectively, and $D_{i} = \partial_{i} - i e A_{i}$ is the covariant derivative.
The fields $s(\Br)$ and $d(\Br)$ can be thought of as $\Delta_{s/s^\pm}(\Br)$ and $\Delta_{d_{xy}}(\Br)$ after coarse graining.
This type of admixing near a vortex has previously been studied in the context of cuprates, 
leading to the prediction of a fourfold-anisotropic order parameter around a vortex \cite{joynt:1990,berlinsky:1995,yong:1995,ichioka:1996a}\footnote{The microscopic model we consider is related to the single-band model of cuprates through rotation by 45$^\circ$, the roles played by $s$-wave and $d$-wave order parameters are reversed and our $d$-wave order parameter is of $d_{xy}$ form rather than $d_{x^2-y^2}$.}.
As the large halo around vortices in cuprates\cite{hoffman:2002} hindered the observation of this admixing, LiFeAs presents an opportunity for this observation.

The spatial variation of the secondary component $d_{xy}$ in Fig.~\ref{fig:gapmap}(c) is largely due to the derivative coupling, the term proportional to $\gamma_v$ in Eq.~\eqref{eq:glfe}.
This intermixing term is expected to be large when the $s$-wave order parameter is of $s^\pm$ type, since the same NNN pairing interaction is reponsible for both $s$-wave and $d$-wave order parameter.
For $|s| \gg |d|$ and $|\vec{D} s| \gg |\vec{D} d|$ the spatial structure of the $d_{xy}$ component is determined largely by the structure of the $s$-wave component.
Minimizing Eq.~\eqref{eq:glfe} with respect to $d(\Br)$ and keeping only terms up to linear order in $d(\Br)$, we find
\begin{align}
-\gamma_{d} \vec{D}^{2} d+\alpha_{d} d + \beta_{3} |s|^2 d + \beta_{4} s^2 d^{*}
= &
 \gamma_{v} ( D_{x} D_{y} + D_{y} D_{x} ) s.
\label{eq:deq}
\end{align}
Hence, the curvature in the leading $s$-wave component will induce the secondary ($d_{xy}$) component.
Now, consider a single isolated vortex. As $s(\Br)$ is recovered at the length scale of the coherence length $\xi$ away from the core of the vortex, we expect a large $d(\Br)$ due to coupling to the large curvature of $s(\Br)$ at this distance. 
Since $\xi = \hbar v_{F} / \pi \Delta \sim 3.0 a_0$ for the uniform gap value with $V'=-0.3 \mathrm{eV}$, this is in agreement with the positions of the maxima of $d(\Br)$ in Fig.~\ref{fig:gapmap}(c) as a function of $|\Br|$ setting the vortex core at the origin.
We can also explain the angular variation and the form of the anisotropy of $d(\Br)$ in this framework.
If we assume $s(\Br) = f(r) e^{i \theta}$ with a slowly changing $f(r)$ and the azimuthal angle $\theta$ measured from the Fe-Fe direction, we find from Eq.~\eqref{eq:deq}
\begin{align}
d(\Br) & 
\sim \partial_{x} \partial_{y}  s(\Br) 
\sim e^{-i \theta} (1 + 3 e^{4 i \theta}),
\end{align}
ignoring the phase due to the magnetic field.
The structure of the derivative hence gives rise to a four-fold anisotropy, which explains the fact that $|d(\Br)|$ is maximum in the Fe-Fe direction, while it is suppressed along the 45$^{\circ}$ direction.
Coupling to $d(\Br)$ gives then in turn cause for the four-fold anisotropy in $s(\Br)$.

\section{Conclusion}

We have contrasted the effects of anisotropic $s^\pm$-wave (NNN) pairing and  isotropic $s$-wave (on-site) pairing on 
 the near-vortex local density of states in LiFeAs by solving Bogoliubov-de Gennes equations both non-self-consistently and self-consistently. 
We have found qualitative changes in the geometric distribution of the density of states as a function of energy.
At low energies, the anisotropy of the vortex bound state, and hence the LDOS, is determined by the normal state low energy electronic structure, independent of the gap structure.
Different pairing structures, however, lead to qualitatively different LDOS distributions at higher energies:
While the isotropic $s$-wave shows a square-like feature of roughly equal intensity, four `hot spots' develop in the case of an (anisotropic) $s^{\pm}$-wave gap. Indeed, our results for the latter case qualitatively agree with recent experiments\cite{hanaguri:2012}.

From the self-consistent treatment we have further found a difference in the recovery of the order parameter away from the vortex core: a pronounced angular dependence of the $s^{\pm}$-wave gap compared to isotropic behavior for the $s$-wave gap.
Employing a Ginzburg-Landau analysis, we have explained this difference through admixing of a secondary order parameter supported by the NNN interaction.
Note that such intermixing is negligible for an $s$-wave pairing with a dominant on-site pairing interaction, as no other pairing instabilities are nearby.
For the NNN interaction, however, $s^\pm$- and $d_{xy}$-wave instabilities have comparable transition temperatures. 
Detection of the anisotropy or even the secondary order parameter would be a strong proof of the unconventional nature of the pairing. 

In this work, we focused on the $\gamma$ band with interest in low energy properties, as this is the band with the smallest gap\cite{borisenko:2010, umezawa:2012}.
Hence, for features at energies less than the gap scale, we expect our calculation to capture salient features of in-field STS experiments.
The comparison between the calculated LDOS for the single- and the five-band models and the results in Ref.~\cite{hanaguri:2012} supports this conjecture. 

In closing we note that our calculation captures Friedel-like oscillations, frequently referred to as quasi-particle interference (QPI), due to vortices.
QPI in the presence of vortices was successfully used  to access phase information with STS in cuprates\cite{hanaguri:2009}.
Recent in-field QPI experiments on FeSe have been interpreted to be consistent with an $s^\pm$ scenario when a vortex is treated as a magnetic scatterer for BdG quasiparticles\cite{hanaguri:2010}.
However, a vortex is at once a point of gap suppression, a point with magnetic flux, and a point around which the order-parameter phase winds.
While we treated vortices faithfully in the self-consistent calculation, we could not investigate effects of inter-pocket sign change as we only considered one pocket.
An extension of the present work with the full band structure would be necessary to work out what to expect for different order-parameter possibilities, especially how the phase difference between different pockets affects in-field QPI.

\ack
We are grateful to M. Allan, J. Berlinsky, J.C. Davis, I. Firmo, T. Hanaguri, C. Kallin,  A.W. Rost,  and Z. Tesanovic for helpful discussions. We acknowledge support from NSF Grant DMR-0955822. MHF and E-AK additionally acknowledge support from NSF Grant DMR-1120296 to the Cornell Center for Materials Research.

\section*{References}

\end{document}